\documentclass[12pt,preprint]{aastex}

\long\def\***#1{{\scshape ***#1***}}



\newcommand{\etal}{{\em et al.}}            

\shorttitle{Cen A discontinuity}
\shortauthors{Kraft \etal}

\begin{document}

\title{Jet Heated X-ray Filament in the Centaurus A Northern Middle Radio Lobe}
\author{R. P. Kraft\altaffilmark{1}, W. R. Forman\altaffilmark{1},M. J. Hardcastle\altaffilmark{2}, M. Birkinshaw\altaffilmark{3}, J. H. Croston\altaffilmark{2},C. Jones\altaffilmark{1}, P. E. J. Nulsen\altaffilmark{1}, D. M. Worrall\altaffilmark{3}, S. S. Murray\altaffilmark{1}}

\altaffiltext{1}{Harvard/Smithsonian Center for Astrophysics, 60 Garden St., MS-67, Cambridge, MA 02138}
\altaffiltext{2}{University of Hertfordshire, School of Physics, Astronomy, and Mathematics, Hatfield AL10 9AB, UK}
\altaffiltext{3}{University of Bristol, Department of Physics, Tyndall Avenue, Bristol BS8 ITL, UK}

\begin{abstract}

We present results from a 40 ks {\em XMM-Newton} observation of the X-ray filament
coincident with the southeast edge of the Centaurus A Northern Middle Radio Lobe (NML).
We find that the X-ray filament consists of five spatially resolved X-ray knots
embedded in a continuous diffuse bridge.  The spectrum of each
knot is well fitted by a thermal model with temperatures ranging from 0.3-0.7 keV and subsolar
elemental abundances.  In four of the five knots, non-thermal models are a poor fit to the spectra, 
conclusively ruling out synchrotron or IC/CMB mechanisms for their emission.
The internal pressures of the knots exceed that of the ambient ISM or the equipartition pressure of
the NML by more than an order of magnitude, demonstrating that they must be short lived ($\sim$3$\times$10$^6$ yrs).  
Based on energetic arguments, it is implausible that these knots have been ionized
by the beamed flux from the active galactic nucleus of Cen A or that they have been
shock-heated by supersonic inflation of the NML.  In our view, the most viable scenario
for the origin of the X-ray knots is that they are the result of cold gas
shock heated by a direct interaction with the jet.  
The lifetimes of the X-ray knots in the NML are roughly the same as the age of the
strong shock around the SW inner lobe, suggesting they were created in the same AGN outburst.
The most plausible model of the NML is that it is
a bubble from a previous outburst that is being re-energized by the current outburst.
We also report the discovery of a large scale (at least 35 kpc radius) gas halo around Cen A.

\end{abstract}


\section{Introduction}

Twin jets launched from the active nuclei of radio galaxies provide an efficient mechanism
for transferring gravitational energy of gas infalling onto a supermassive black hole (SMBH)
back into thermal energy of the ambient gas out to distances hundreds of kpc from the nuclei.
The conventional paradigm suggests that the synchrotron-emitting radio lobes of radio
galaxies contain relativistic plasma supplied by jets.
In the early stage, the momentum of the jets dominates the dynamical evolution.  
Depending on the mechanical power of the jet and richness of the ambient environment,
the jet will initially drive a strong shock into the ambient gas.
As the bubble becomes large and the radio galaxy evolves, the advance speed of the jet head decreases
and the energy/pressure of the bubble determines the dynamical evolution.
Once the energy supply from the active nucleus ceases, the bubble will come into rough
pressure equilibrium with the ambient gas and its dynamics will largely be driven by
buoyancy in the stratified atmosphere.

Centaurus A is the nearest radio galaxy to the Milky Way ($d$=3.7 Mpc, 1$'$=1.076 kpc), and exhibits a
complex morphology on spatial scales from milliarcseconds to degrees.
A montage of the radio emission from Cen A on various spatial scales is shown in Figure~1 of
\citet{morg99}.  VLBI observations of Cen A (spatial scales of tens of
milliarcseconds) detected the bright AGN, a forward jet,
and a counter-jet.  Based on measurements of the sidedness of the VLBI jets, the jet velocity
is estimated to be $\sim$0.3$c$.
On arcminute scales, \citet{burns83} detected a one-sided radio jet extending $\sim$4$'$ to
the northeast of the nucleus opening into a radio lobe (the NE inner lobe, or simply NE lobe).  
A second lobe of similar size and radio brightness was detected to the southwest (the SW lobe) with
no obvious radio counter-jet feeding it.  \citet{burns83} found only a few spatially resolved radio knots embedded
in the plasma of the SW lobe.

{\em Chandra} and {\em XMM-Newton} observations of the inner jet and radio lobes show significant differences
between the NE lobe and the SW lobe.  The dominant X-ray emission mechanism from the
inner jet is synchrotron radiation emanating from $\sim$30 knots embedded in diffuse emission.  
There are significant spatial offsets between
the X-ray and radio peaks of the inner jet \citep{kraft02}, and proper motions of some radio knots show
velocities of $\sim$0.5$c$ \citep{mjh03}.  Deep {\em Chandra} observations find extended 
diffuse emission associated with the radio knots in the SW lobe, conclusively 
demonstrating that this is the counter jet \citep{mjh07}.
A shell of shock-heated gas and X-ray synchrotron emitting ultra-relativistic
electrons surrounds the SW lobe \citep{kraft03,kraft07,croston09}.  The temperature and density of the gas in the shell
is several times that of the ambient medium.  Thus, the inflation of the SW lobe is driving
a strong shock into the ISM.  Surprisingly, there is no corresponding shell of shock-heated gas
around the NE lobe.  There are a few filaments of X-ray emission around the NE lobe that may or
may not be shock-heated gas, but the emissivity of these features is roughly two orders of magnitudes 
lower than the shell around the SW lobe \citep{kraft07}.

A much larger, lower surface brightness radio lobe, called the Northern Middle Lobe (NML),
lies to the northeast beyond the NE (inner) lobe, $\sim$20-30 kpc (roughly 20$'$-30$'$)
from the active nucleus \citep{junk93}.
This lobe is connected to the NE lobe by the large scale jet (LSJ) \citep{morg99}.
Whether the LSJ is actually a jet in the traditional sense that it represents a
continuous channel for the transfer of energy and momentum in a collimated flow is unclear based
on its radio morphology.
Several sets of optical emission line filaments are associated with the NML and LSJ.
The inner optical filament lies $\sim$2$'$ to the southeast of the LSJ \citep{blanco75} and is
coincident with weak X-ray emission \citep{evans04}.  The outer filaments lie in the interior
of the NML in projection, roughly 15-20$'$ from the nucleus \citep{graham98}.  Several regions
of star formation also lie in the interior of the NML in projection, perhaps the best known
examples of jet-induced star formation \citep{mould00}.
An HI cloud lies at the base of the NML just north of where the LSJ enters
the lobe \citep{oost05}.  Interestingly, there is no radio feature corresponding to the NML 
lying beyond the SW (inner) lobe.
It has been proposed that the NML is either a buoyant bubble from
a previous epoch of nuclear activity \citep{sax03} or that the NE inner radio lobe has burst \citep{morg99}.
In the first scenario, the age of the NML would be on the order of several hundred Myrs, in the
latter its age would be roughly the age of the inner radio lobes - only several Myrs.

\citet{feig81} reported the discovery of an X-ray filament along the southeast edge
of the NML from an {\it Einstein} IPC observation.  
This feature was also detected in EXOSAT \citep{morini89}, {\em ASCA} \citep{morg99}, and
an unpublished archival {\em ROSAT} PSPC observation.
\citet{feig81} argued that this feature was probably from hot gas, and rejected both
synchrotron emission from a population of ultra-relativistic electrons and inverse-Compton 
scattering of CMB photons as viable possibilities.  
The synchrotron scenario would require an unusual X-ray to radio flux ratio, very different
from that seen in FR I jets in which the X-ray emission is believed to be synchrotron.
Likewise, the inverse-Compton scattering of the microwave background
photons (IC/CMB) hypothesis would require a large ($B_{NML}<B_{eq}/10$) departure from 
equipartition, much larger than is seen in a sample of radio lobes observed with {\em Chandra}
and {\em XMM-Newton} \citep{cro05}.
In the IC/CMB scenario, the morphology of the X-ray emission must match the distribution
of the low energy electrons and therefore be somewhat similar to the radio lobe (which depends
on both the electron distribution and magnetic field strength).  The morphological
dissimilarities between the emission in the two bands offer a second strong argument
against the IC/CMB scenario.
The X-ray filament is associated with a region of lower radio polarization in the NML,
adding support to the thermal gas hypothesis \citep{morg99}.  
\citet{sax03} have modeled the NML as a buoyant bubble and identify the X-ray filament
as the thermal gas trunk of material that has been drawn up behind the bubble from the
central regions of the host galaxy.

Optically Cen A is crossed by a dark dust band that is believed to be the result
of a merger with a small spiral galaxy similar to M33
several hundred million years ago \citep{malin83,israel98}.
We note that there is considerable debate in the literature as to the age of the merger,
with low estimates on the order of 100 Myrs \citep{quil92} and high estimates
of several Gyrs \citep{soria96}.  The low end of this range is roughly consistent with
the age of the NML if it is a buoyant bubble.
The merger certainly induced large scale gas motions that, depending on the time since the merger and
the viscosity of the ISM, may or may not have dissipated.  Such gas motions  may play an important role
in shaping the large scale morphology of the radio source.
It is possible that the accretion of cold gas by the central SMBH provided by the spiral is
fueling the current epoch of nuclear activity.
The merger may well have fueled the earlier epochs of nuclear activity that created the radio
structures on tens of arcminutes and degree scales.

In this paper, we present results from a 40 ks {\em XMM-Newton} observation of the
X-ray filament associated with the Cen A NML.
The primary science objective of this observation was to determine the nature
of the X-ray emission from the NML filament and to determine whether the
NML is produced by the current nuclear outburst or is a remnant of
an earlier one.  This paper is organized as follows.  Section 2 contains a brief summary of the data.
Analysis is presented in section 3, followed by an interpretation of our results
in Section 4.  Section 5 contains a brief summary and conclusion.
We adopt a distance of 3.7 Mpc to Cen A, which is the mean of distance measures
from five independent indicators (TRGB, Mira variables, SBF, PNLF, Cepheids,
summarized in Section 6 of \citet{fer07}).
At this distance, 1$''$=17.9 pc and 1$'$=1.076 kpc.
All uncertainties are at 90\% confidence for one parameter of interest
unless otherwise stated, and all coordinates are J2000.
All spectral fits include absorption ($N_H$=8.41$\times$10$^{20}$ cm$^{-2}$)
by foreground gas in our Galaxy \citep{dic90}.

\section{Data Analysis}

The Cen A NML was observed for $\sim$40 ks with {\em XMM-Newton} on February 2, 2006 (OBSID 0302070101,
revolution 1127) with the Medium filter inserted.  The MOS and PN event files were filtered using a strict
criterion for highest energy resolution.  All events in the MOS event files with FLAG$>$0 and
PATTERN$>$12 were excluded.  All events in the PN event file with FLAG$>$0 and
PATTERN$>$4 were excluded.  Light curves were made for all three event files in the
5.0-10.0 keV band and periods when the background exceeded the mean value by more than 3$\sigma$
were removed.  The time removed due to background flaring was relatively small.
Flare filtering left 38477 s, 38126 s, and 26540 s of good time in the MOS1, MOS2, and PN
cameras, respectively.  All point sources visible by eye were excluded from our analysis.

Data from three other X-ray observations are used in the analysis presented in this paper.
The central region of Cen A was observed with {\em XMM-Newton} twice (OBSIDs 0093650201 and 
0093650301) for observation times of 36055 s, 36045 s, and 24758 s in the MOS1, MOS2
and PN cameras, respectively.  Results from these observations were originally
published in \citet{kraft03} and \citet{devans04}.  The southern part of the Cen A NML is also contained 
within the FOV of one {\em Chandra}/ACIS-I observation (OBSID 7800 - 100 ks).  Several of the
knots of the NML are located on the S2 chip in this observation, $\sim$15$'$ off axis.
Since the NML is so far off axis, we use these data only to confirm the measured temperatures
of several of the fainter knots (N3, N4, and N5) at the southwestern boundary of the filament.
Data from an archival {\em ROSAT} PSPC observation ($\sim$13.5 ks) is used to constrain the nature of the
large scale gas halo.
Finally, we use previously published VLA \citep{burns83,mjh03} and ATCA \citep{morg99} radio data taken from 
public archives for comparison with our X-ray data.

\section{Analysis}

\begin{figure}
\plotone{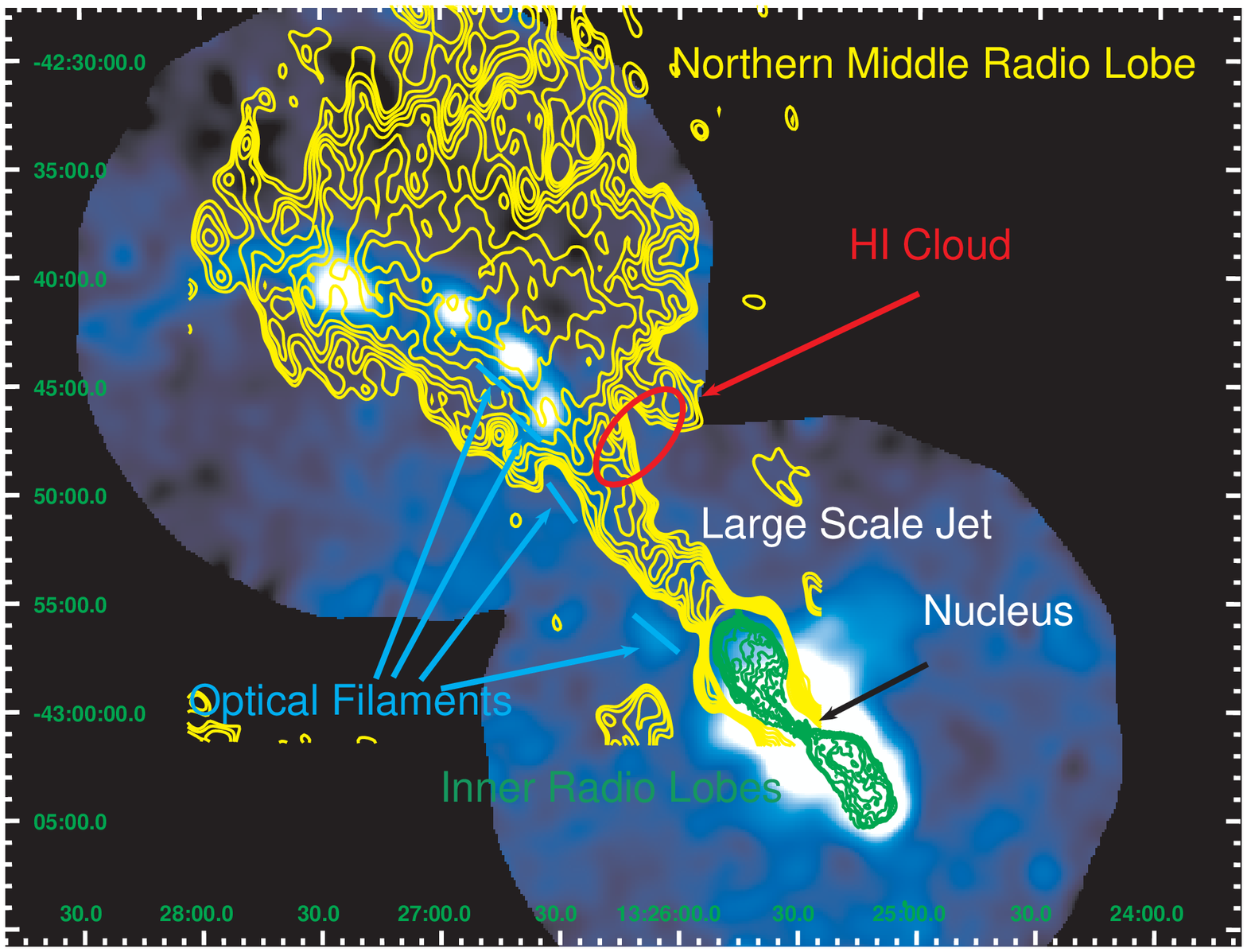}
\caption{Exposure corrected, Gaussian smoothed {\em XMM-Newton} image of the inner jet/lobe
and NML regions of Cen A in the 0.3-1.0 keV band pass.  The inner jet and inner radio
lobes are shown with the green contours \citep{burns83}, the NML as the yellow
contours \citep{morg99}.  The red circle denotes the approximate
position of the HI cloud located at the base of the NML \citep{sch94,oost05}.
The position of the active nucleus is shown with the black arrow.
The approximate positions of the inner and outer optical filaments \citep{blanco75,morg99}
are shown as the cyan lines.}\label{nmlovl}
\end{figure}

The large-scale spatial relationship between the X-ray filament of the NML, the radio components of the NML
and inner lobes, and the X-ray gas of the central regions of Cen A is shown in Figure~\ref{nmlovl}.
This is a Gaussian smoothed, exposure corrected MOS (1+2) image in the 0.3-1.0 keV band 
of all the {\em XMM-Newton} data on Cen A.
The inner jet and inner radio lobes are shown by the green contours, the NML with the yellow
contours.  The X-ray filament lies along the southeastern boundary of the NML.  Approximate
positions are indicated for several other features of interest,
including the HI cloud (red) and the inner and outer optical filaments (cyan) are also shown.
There are clear peaks in the radio emission between the pairs of X-ray knots N1/N2, N2/N3, and N3/N4.
The same image is shown in Figure~\ref{nmlblue} without the radio contours or other overlays
to more clearly distinguish the X-ray filament, the large scale diffuse emission (the light blue),
and the background level (the black).

An unsmoothed X-ray image (MOS1+MOS2 co-added, 4$''$ per pixel) of the Cen A NML in the 0.5-2.0 keV
band is shown in Figure~\ref{nmlraw}.  The X-ray filament of the NML is composed of five distinct
X-ray knots of emission (labeled N1 through N5 in Figure~\ref{nmlraw}) with a bridge of diffuse
emission connecting all the knots and extending to the east beyond knot N1.
The point sources have been retained in this image, clearly showing that all five knots are spatially
extended at the resolution of {\em XMM-Newton}.
Figure~\ref{nmlsmooth} contains a smoothed (Gaussian - $\sigma$=24$''$ kernel), 
exposure corrected (MOS1+MOS2 coadded) image in the 0.5-2.0 keV band of the NML X-ray filament.
The three knots closest to the nucleus (N3, N4, and N5) are also contained within the S2 chip of an ACIS-I observation
of Cen A as shown in Figure~\ref{nmlchandra}.  The NML is $\sim$15$'$ off-axis in this observation, so
the PSF is fairly large, but the three knots can clearly be seen, as well as the diffuse emission between
the knots.  Another diffuse feature lies 2.5$'$ to the NW of knot N3, perhaps related to the rest
of the filament.  Two bright point sources are located at the SW edge of the chip, indicative of
the size of the PSF in this observation.
Interestingly, the X-ray knots are anti-coincident with the radio knots of the NML.  This can
be seen most clearly in the radio map with X-ray contours overlaid in Figure~\ref{nmlgap}.

We fitted thermal and power-law models to the
spectra of each of the five knots.  Circular extraction regions were used
to include all of the emission for each of the five knots.  The positions and
sizes of these regions are given in Table~\ref{specfit}.  Background was determined from
two different local regions, one just to the northwest of the X-ray filament and a second along
the periphery of the field of view.  The results are insensitive to which background region is
used in the fits.  We quote results using the first background region only.
Data from the MOS1, MOS2, and PN cameras were fitted simultaneously.  

For each knot, the spectra were fitted with two models:
an absorbed power law spectrum and an absorbed thermal (APEC) model with the elemental
abundance as a free parameter.  In all cases the absorption
was held fixed at the Galactic value.  Allowing this parameter to vary freely made no statistically
significant difference in any of the fits, and the statistical uncertainty in this value
was always consistent with the Galactic value.  For knots N1 through N4, the power law model
can be ruled out at high ($>>$99\%) confidence, thus strongly disfavoring a non-thermal
(synchrotron or IC/CMB) origin for the X-ray emission.  In all cases the thermal model
provided an acceptable fit to the spectrum.  A summary of the best fit temperatures
and abundances is shown in Table~\ref{specfit}.

\subsection{The Structure of Knot N1}

There is considerable substructure present in knot N1 which can be seen more
clearly in Figure~\ref{n1fig}.  The bridge structure connecting N1 to N2 lies
along the east/west axis and extends beyond N1 to the east.  
The peak of the X-ray emission lies roughly 1.2$'$ south of where this bridge
runs through the knot.  In fact, the emission from all of the knots generally lies to the south and east
of the thin bridge that connects them.  The bridge extends to the east of knot N1 past the edge of the
radio lobe (see Figure~\ref{nmlovl}).  This suggests that the radio emission from the NML 
visible in the ATCA 1.5 GHz maps does not define the full extent of the NML.

\subsection{Knot N5}

The best fit temperature of knot N5 is considerably higher ($\sim$1 keV in the {\em XMM-Newton}
data, and $>$2.5 keV in the {\em Chandra} data) than that of the other four knots. 
In our favored model described below, in which the X-ray knots represent jet-cloud interactions,
the higher temperature of N5 could suggest that the ram pressure of the jet
impacting this cloud is somewhat higher than further along into the NML.
It is also possible, however, that this feature is somehow different from the others or even unrelated to
the NML.  Knot N5 is clearly extended in both data sets, and is connected to 
knot N4 by the bridge, supporting the idea that this feature is part of the X-ray structure associated
with the NML.  On the other hand, its spectrum can be well fitted by an absorbed power-law
model (photon index 1.8$\pm$0.3), consistent with the synchrotron or IC/CMB emission mechanisms.
We proceed with the assumption that this feature is thermal gas related to the rest of the X-ray
filament.  Our results and general conclusions on the X-ray features as a whole are unaffected if
some of all of the flux from this feature is due to non-thermal emission or
(perhaps) unresolved point sources.

\subsection{Bridge structure connecting knots}

We attempted to fit the {\em XMM-Newton} spectra of the thin bridge connecting the 5 bright X-ray knots,
but the count rate was too low and the background too high to obtain a meaningful result.  We did fit
the emission between the pairs N3/N4 and N4/N5 in the {\em Chandra} data using the off-axis
S2 CCD.  The data were combined
into one spectrum to maximize signal to noise, and a local region on the S2 chip
away from the bridge was used to estimate background.  The spectrum is well described by 
an absorbed thermal model with a temperature of 0.72$\pm$0.06 keV. The abundance is poorly constrained.  
A single power-law model can be rejected at high significance.  The spectrum is entirely consistent with that
of the bright knots confirming that they are of similar origin.

\subsection{Hot ISM of Cen A}

\begin{figure}
\plotone{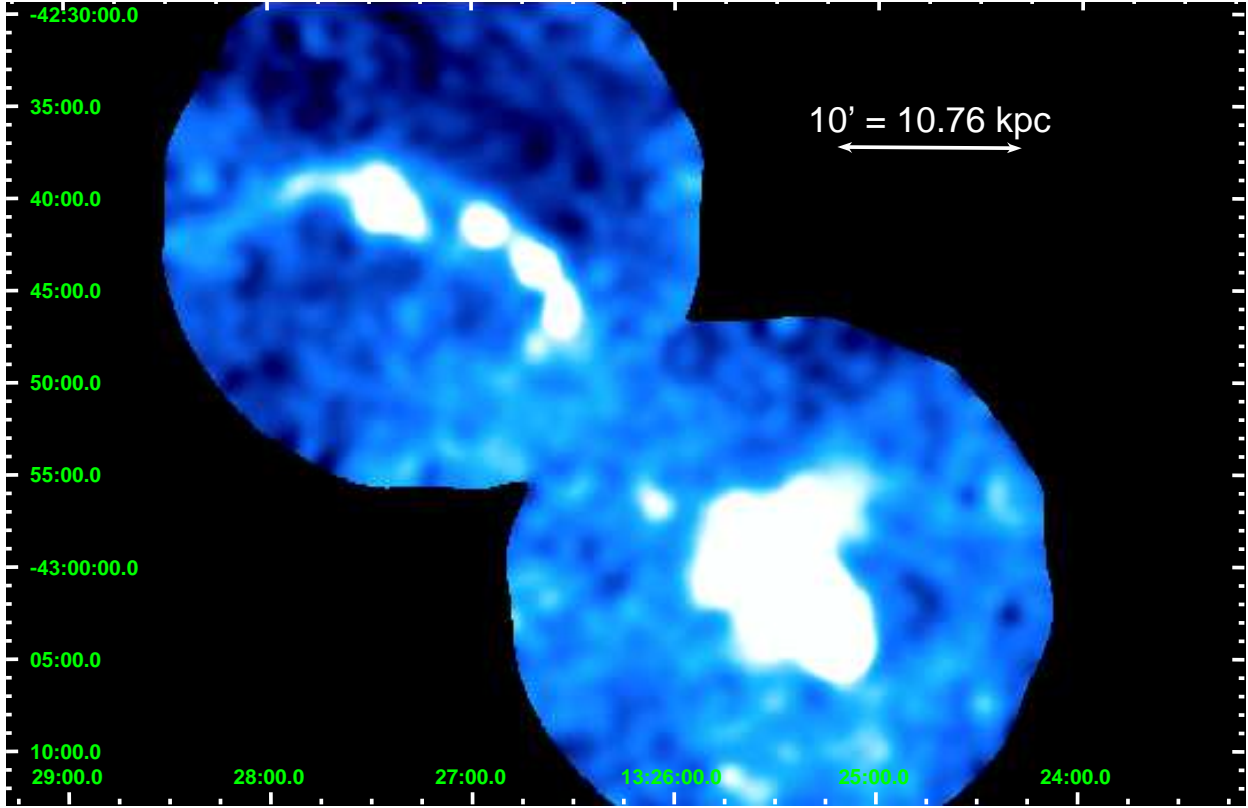}
\caption{Exposure corrected, Gaussian smoothed {\em XMM-Newton} image of the inner jet/lobe
and NML regions of Cen A in the 0.3-1.0 keV band pass.  
This is the same image as shown in the previous figure without the radio contours and
a slightly different color stretch to enhance the appearance of the large scale diffuse emission.
Notice the surface brightness contrast to the NW and SE of the X-ray filament.  The light blue
and black (the level of the sky background)
colors correspond to surface brightness of approximately 3.2 and 5.2$\times$10$^{7}$
cts s$^{-1}$ arcsec$^{-2}$.}\label{nmlblue}
\end{figure}

\begin{figure}
\plotone{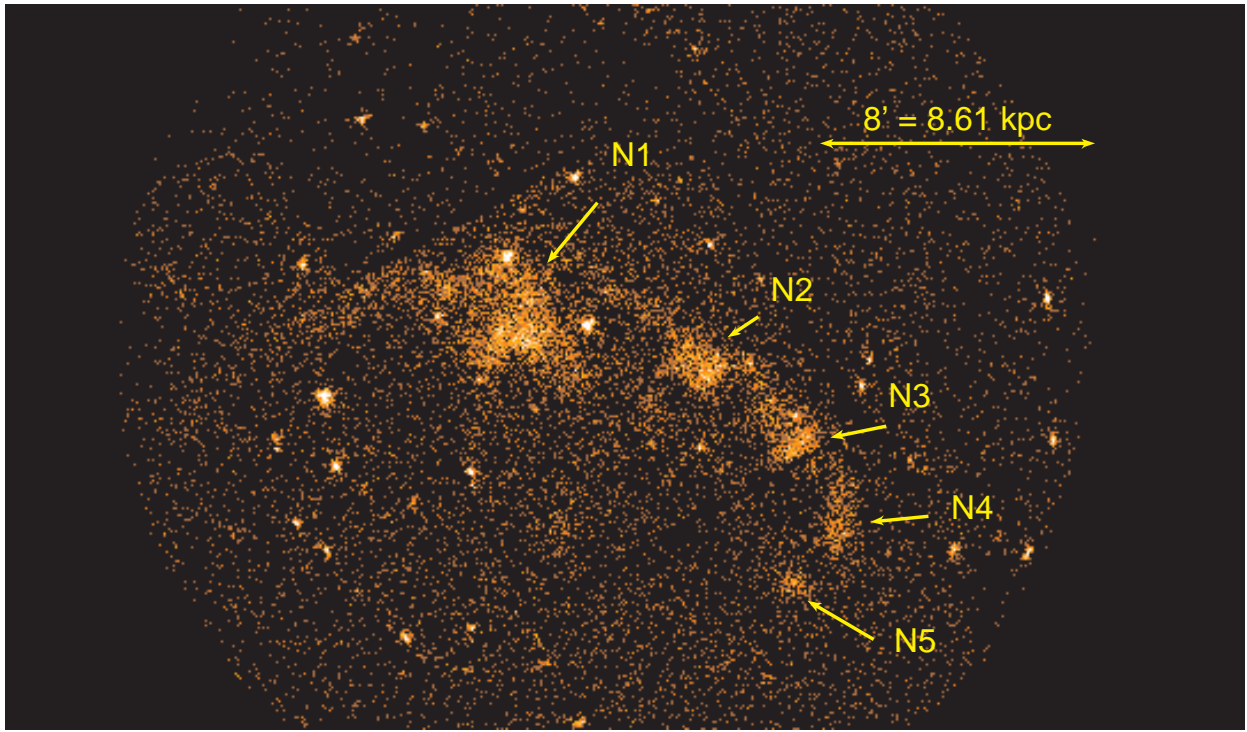}
\hspace{0.5in}
\caption{Unsmoothed {\em XMM-Newton} image of the Cen A NML (MOS1+MOS2 co-added) in the 0.5-2.0
keV band.  The five X-ray knots are labeled N1 through N5.  All five are spatially
extended at the resolution of {\em XMM-Newton}.}\label{nmlraw}
\end{figure}

\begin{figure}
\plotone{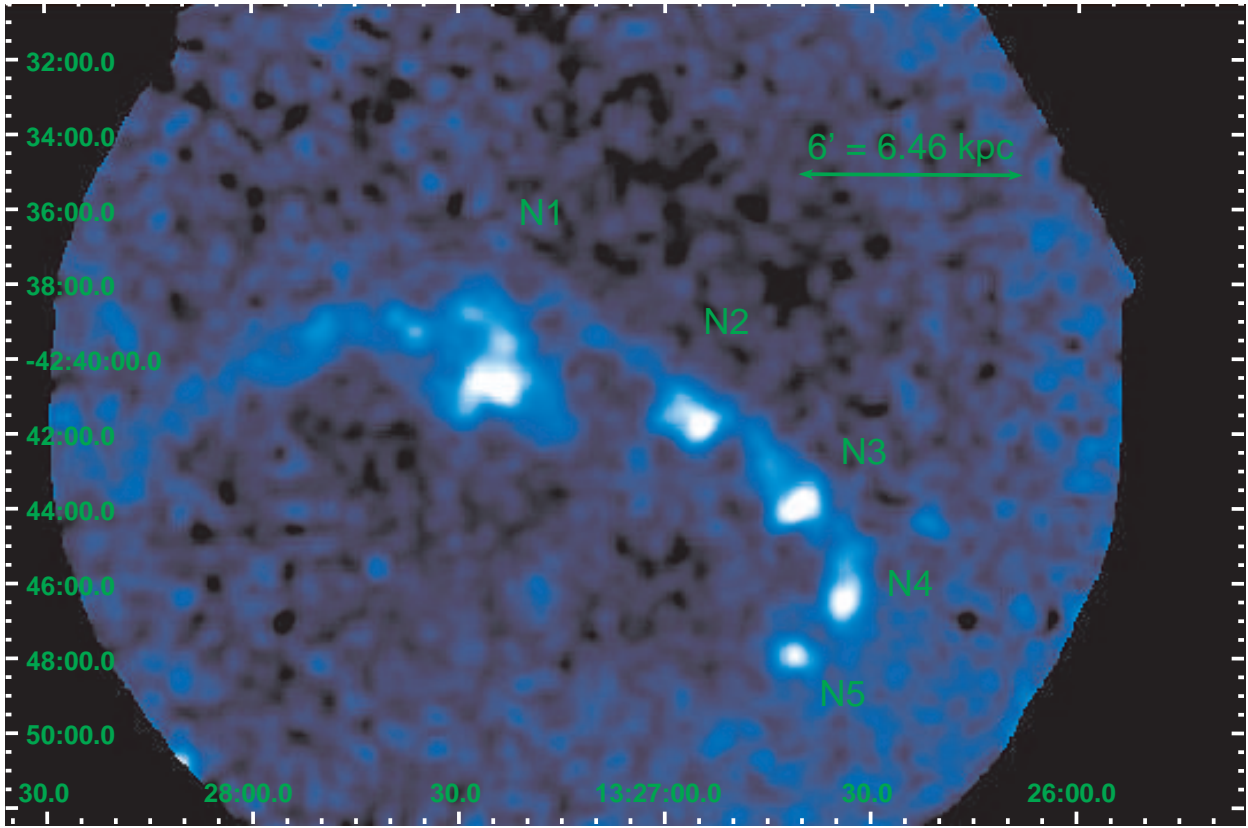}
\hspace{0.5in}
\caption{Smoothed (Gaussian kernel - $\sigma$=24$''$), exposure corrected
{\em XMM-Newton} MOS1+MOS2 image of the Cen A NML in the 0.5-2.0
keV band with point sources removed.}\label{nmlsmooth}
\end{figure}

\begin{figure}
\plotone{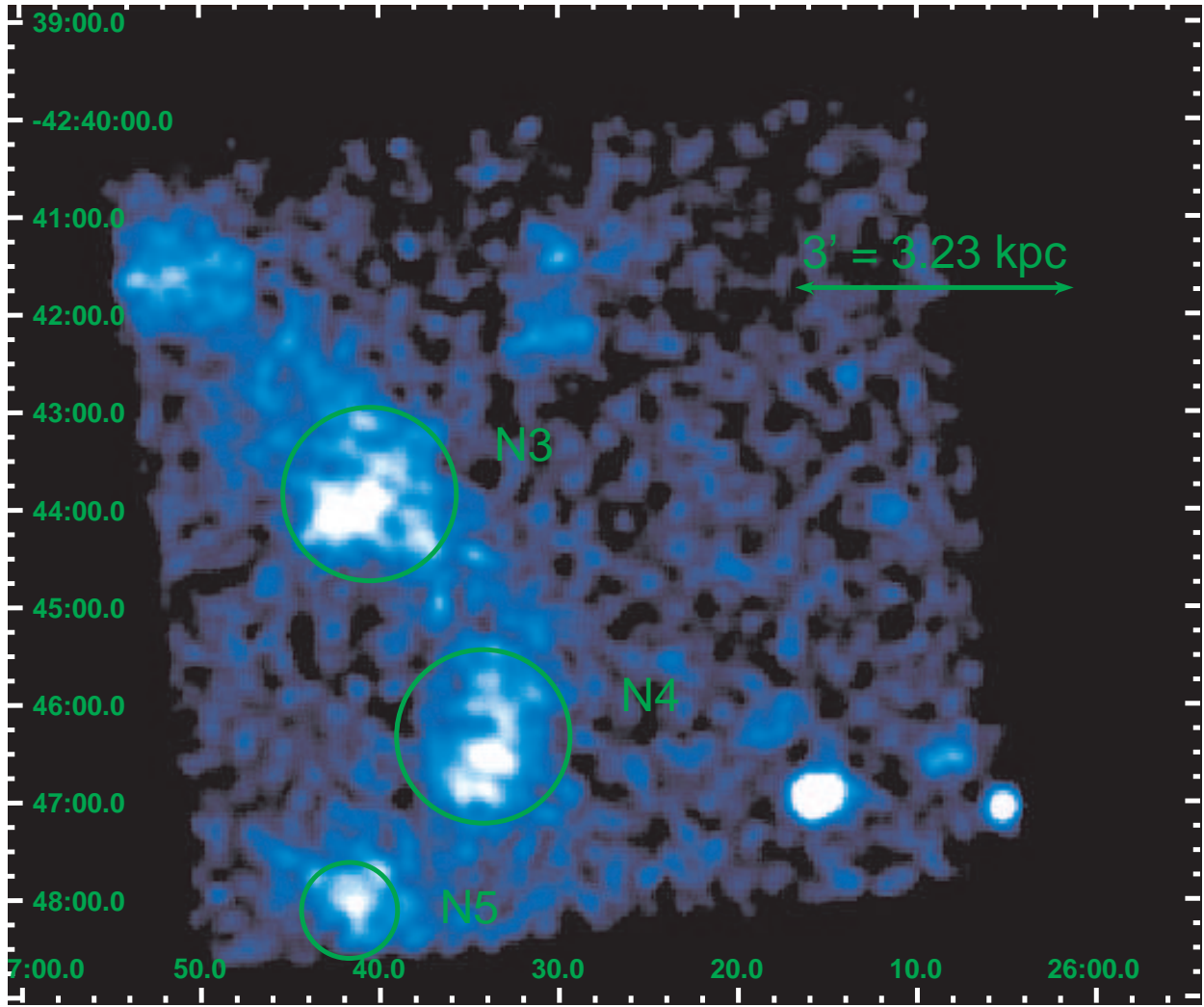}
\hspace{0.5in}
\caption{{\em Chandra} image of knots N3, N4, and N5 of the Cen A NML in the 0.5-2.0
keV bandpass.  This region lies roughly $\sim$15$'$ from best focus.  Two point sources
to the west of the filament can be used to visually estimate the PSF.  All the knots
are clearly extend.}\label{nmlchandra}
\end{figure}

\begin{figure}
\plotone{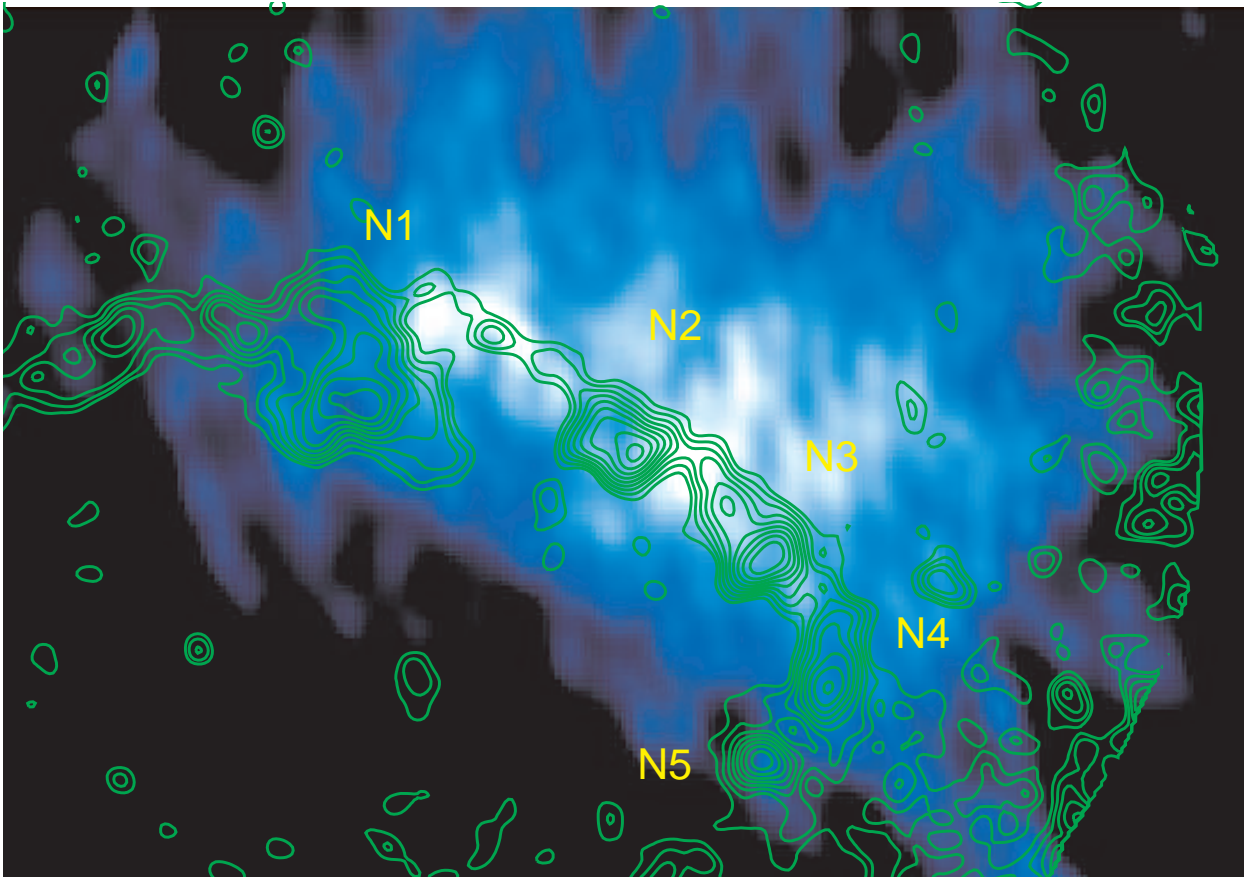}
\hspace{0.5in}
\caption{Radio map of Cen A NML (1.34 GHz, 60$''$(NS)$\times$36$''$(EW) beam) with X-ray contours from
Figure~\ref{nmlsmooth} overlaid.  Note the anti-coincidence between the X-ray and radio
peaks around knots N1, N2, and N3.}\label{nmlgap}
\end{figure}

\begin{figure}
\plotone{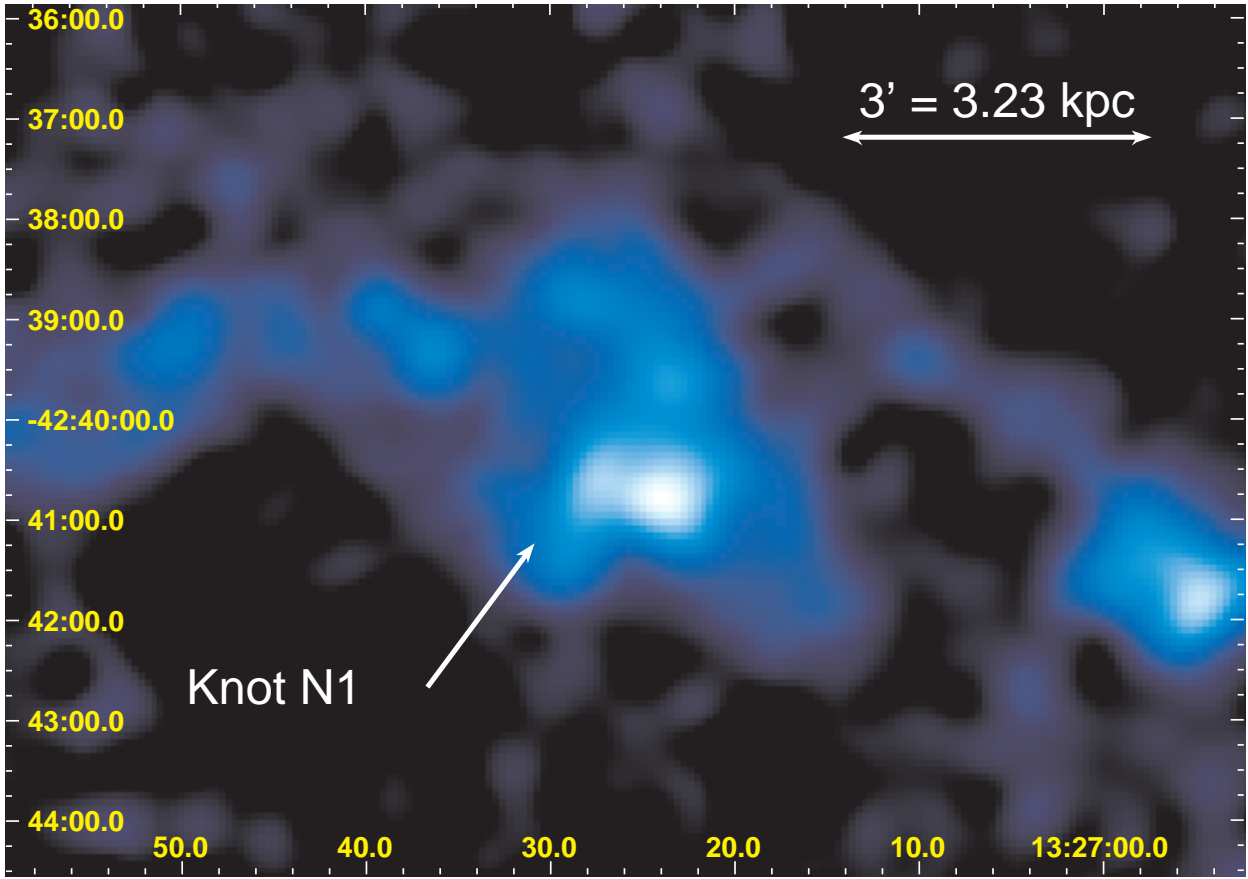}
\hspace{0.5in}
\caption{{\em XMM-Newton} MOS1+MOS2 image (exposure corrected with point sources
removed) of knot N1 of the Cen A NML in the 0.5-2.0
keV band.}\label{n1fig}
\end{figure}

There is clearly an asymmetry in the diffuse X-ray emission on either side (NW/SE) of the X-ray filament
(see Figures~\ref{nmlovl} and~\ref{nmlblue}).  We interpret this excess diffuse emission to the SE of the filament
as the hot ISM of Cen A.  There is a region of enhanced X-ray surface brightness (the light blue
in Figure~\ref{nmlblue}) that extends more or less continuously toward the nucleus.  This is the
extended gas corona of Cen A.  In fact, diffuse X-ray emission fills the XMM-Newton FOV of both
observation and therefore extends more than 30 kpc (30$'$) from the active nucleus.
The actual extent of the ISM is unclear, an observation over a larger FOV is required to map its
extent.  We note that the cavity filled by the radio plasma of the NML
appears to be devoid of this X-ray emission (the black in Figure~\ref{nmlblue}).
The inflation of the NML must have evacuated this cavity; the lack of X-ray emission
demonstrates that the radio plasma and thermal gas are not well mixed.  The visibility of the
X-ray depression confirms that the lobe lies fairly close to the plane of the sky.  If the
lobe were far from the plane of the sky, the evacuated cavity would not make
a noticeable decrement in the X-ray emission.
There is considerable asymmetric temperature and surface brightness structure in this
emission in both the {\em XMM-Newton} and {\em Chandra} observations.  The general structure of the diffuse
emission is also seen in the {\em ROSAT} PSPC observation of Cen A.  Details of this structure will be
presented in a future publication (Kraft \etal~in preparation).  For the purposes of this paper,
we only need to obtain a rough estimate of the gas density and pressure to assist us in our
interpretation of the emission coincident with the NML.
We note that this extended halo was treated as background in \citet{kraft03}.  Density estimates
in the central regions of the galaxy are still approximately corrected, but any extrapolation of the
surface brightness profiles shown in \citet{kraft03} to more than several arcminutes 
from the nucleus do not account for this extended halo as it was subtracted as background.

\begin{figure}
\plotone{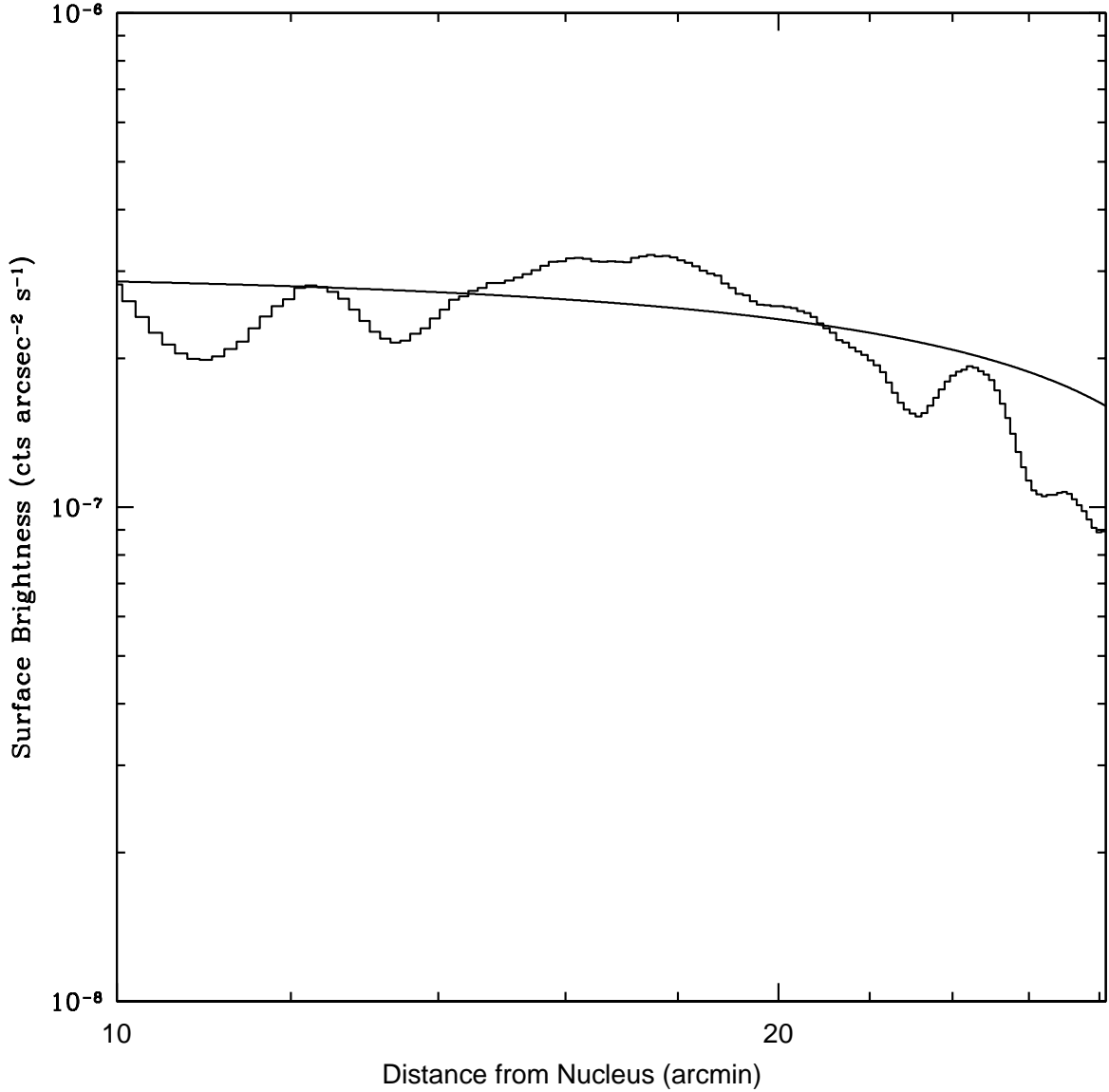}
\hspace{0.5in}
\caption{Surface brightness profile (histogram) of hot gas in a 10$^\circ$ wedge (0.3-1.0 keV) southeast of the
X-ray filament with vertex at the nucleus.  This profile was created from a Gaussian smoothed 
($\sigma$=30$'$ exposure corrected MOS1+MOS2 image.  The continuous curve shows the projected surface brightness
of the uniform density sphere discussed in the text.  The background level (3.6-4.0$\times$10$^{-7}$ cts 
s$^{-1}$ arcsec$^{-2}$ was determined in a region to the northeast of the X-ray filament.}\label{sbprof}
\end{figure}

The surface brightness profile of the gas in a 10$^\circ$ wedge in the 0.3-1.0 keV band
is shown in Figure~\ref{sbprof}.
This profile was taken from a smoothed, exposure corrected
image (MO1+MOS2 combined), and the wedge lies just below (south) of the large scale jet and NML. 
The vertex of the wedge is at the nucleus.
Background was determined from a distant region to the NW of the X-ray filament.  
We assume that the radio lobe completely evacuates a cavity
in the ISM, and that the integrated path length through the gas
outside of the lobe along our line of sight is small.  Some fraction of the emission in this region
may be due to gas in Cen A, but the background level at the interior of the lobe is roughly consistent
with XMM-Newton dark sky background level.  This could make a small difference to our derived density, but
since the density goes as the square root of the surface brightness, the potential error is smaller than
the uncertainty in the density due to the abundance.
We determined the temperature of the hot coronal gas by fitting the spectrum of a circular region
275$''$ in radius at the base of the NML in the {\em XMM-Newton} data.  
We find a temperature of 0.35$\pm$0.04 keV with low abundance.
We model the density profile as a sphere of uniform density with a radius of 35 kpc (the results
are insensitive to this choice).  
This model is obviously only an approximation that facilitates order of magnitude estimates of density
and pressure.  Such a flat profile out to large radii (20 kpc or more) is atypical for isolated
early-type galaxies like Cen A and suggests that the large scale corona is still dynamically
evolving either from the merger or from various epochs of nuclear activity.
Deprojecting the surface brightness profile and using the best fit temperature and 
assuming $Z$=0.5, we find a hydrogen density of $\sim$10$^{-3}$ cm$^{-3}$. 
The surface brightness profile corresponding to this density model is overplotted onto Figure~\ref{sbprof}.
The gas pressure in this sphere is $\sim$8.5$\times$10$^{-13}$ dyne cm$^{-2}$.

The discovery of a large scale gas halo is Cen A is entirely consistent with what
we know about Cen A as the dominant member of a group.  This halo has not been previously reported
due to its angular extent (as much as a degree or more on the sky) and low surface brightness.
Cen A is known to be the dominant member of a poor group with a total gravitating mass of
1-2$\times$10$^{13}$ M$_\odot$ to a distance of $\sim$640 kpc \citep{hesser84,bergh00}.
The total gas mass out to 35 kpc (based on our density model above) is $\sim$3$\times$10$^{9}$ M$_\odot$.
There is likely to be considerable gas out to larger radii.  {\em Chandra} observations of
groups and clusters of galaxies show that baryon fraction is typically a few percent of the
gravitating mass out to $r_{500}$ \citep{finog01,khos07,sun08}.  Therefore the total
gas mass of Cen A is likely to be 2-3$\times$10$^{13}$ M$_\odot$.  Even out to 35 kpc, we
are seeing only a small fraction of the Cen A IGM.  A recent HI survey of Cen A dwarf ellipticals
confirms the presence of the Cen A gas halo.  \citet{bouchard07} found that many Cen A
dwarf galaxies were devoid of HI and suggested that it had been ram pressure stripped by
gas at density on the order of 10$^{-3}$ cm$^{-3}$.  Such an extended halo around Cen A is
entirely consistent with our X-ray observations.

\section{Interpretation}

Based on the spectra, we conclude that the X-ray filament consists of hot gas that lies along
the southeastern boundary of the NML.  
Given the thermal spectra (for four of the five knots), we rule out a non-thermal
(synchrotron or inverse-Compton scattering) origin for the X-ray knots.
The proton density, pressure, mass, thermal
energy, and sound crossing time for each knot (assuming the values from the
best-fit collisional plasma model) are tabulated in Table~\ref{paramtab}.
We have assumed spherical symmetry and unity filling factor for each of the features, and chosen a radius
that roughly corresponds to the largest dimension for each feature.  
We also have assumed that $n_e=1.2n_H$, appropriate for a plasma with sub-Solar
elemental abundances.  The lifetime is the sound crossing time of the knot and is defined as the radius
divided by the sound speed (for $\gamma$=5/3).
A smaller filling factor (i.e. if the gas is clumped and/or if considerable unresolved
structure exists in a given feature) would increase the gas density and pressure of the individual
clumplets, but decrease the total mass, thermal energy, and lifetime of the gas.
The low elemental abundances found in the spectral fits are probably not realistic and likely
the result of unresolved multi-temperature gas.  In such cases, the best fit
temperature usually represents a reasonable average temperature, but the low abundance would imply
an artificially large gas density and pressure.  We have scaled all the XSPEC normalizations, $N$, from the
spectral fits to $Z$=0.5 Solar assuming that $N\propto Z$ to provide more realistic estimates
of gas pressures, densities, and total masses.

The X-ray knots/bridge and associated optical emission features of the Cen A NML are analogous to
similar features seen in the Extended Emission Line Region (EELR) of a large number of 
low redshift radio galaxies as well as many more powerful, distant radio galaxies \citep{baum88}. 
Individual examples of jet-cloud interactions studies with {\em Chandra} include the nearby
3C 277.3 \citep{tadhunter00} and PKS 2153-699 \citep{ly05, young05} as well as the more 
distant (z=0.81) 3C 265 \citep{sol03}.
The EELR region is typically defined as the region $>$1 kpc 
and $<$20 kpc from the nuclei of radio galaxies so that it lies beyond the narrow line region 
(NLR) of the active nucleus but still within the host galaxy.
In fact, the multi-phase gas features and regions of star formation in Cen A lie at somewhat
greater distances from the nucleus (up to 30 kpc) than the EELR region, nominally outside
of the galaxy.  Energy deposition into the ISM via shock-heating by radio outflows
and photoionization from nuclear emission plays a key role in the overall energy budget
of the galaxy and its evolution.  Jet-induced star formation is also a common phenomenon seen 
in both nearby objects, including Cen A \citep{oost05,vanbreugel2004}, 
and high redshift elliptical galaxies in the
process of formation and undergoing their last epoch of significant star formation \citep{dopita2007}.
Such interactions may also be important in the regular refueling of nuclear activity \citep{tad89}.
A general review of this phenomenon is given in \citet{tadhunter02}.

There are {\it a priori} at least five plausible scenarios for the origin of the thermal gas filament
of the Cen A NML:  supernovae-driven winds from regions of jet induced star formation, 
photo-ionization from the beamed X-ray flux from the parsec scale jet,
entrainment of hot ISM from the central region of the galaxy by the NML as it rises buoyantly, 
shock-heating of ambient gas due to the supersonic inflation of the NML, shock-heating
of the ambient gas due to supersonic gas motions of the ISM, and warm/cold gas that is 
being directly heated by the jet.
We discuss each of these possibilities in more detail below and find
serious problems with each of the first five scenarios.  The most likely
possibility is that the X-ray emission originates from warm/cold gas that is shock heated by the jet.

One important constraint on any physical model for the origin of the X-ray
filament is the age of the NML.  As elucidated
below, there are two general possibilities:  either NML is part of the current epoch of
nuclear activity and has the same age as the inner radio lobes (roughly 3 Myrs - see \citet{kraft03}
and \citet{croston09}), or the NML is part of a previous epoch of of nuclear activity unrelated to
the inner radio features and has risen
buoyantly to its present position (age is roughly 100-200 Myrs \citep{sax03}).
We favor a scenario in which the NML was formed as part of a previous epoch of nuclear activity
that has been re-energized by the present outburst.

\subsection{Star Formation - Supernova-Driven Hot Gas Bubbles}

The large thermal energy and mass of the gas argues against the possibility that the X-ray filament 
and knots of the NML are the result of a supernovae-driven wind from a
recent epoch of jet induced star formation.
The total thermal energy of the filament is $\sim$10$^{55-56}$ ergs, requiring
10$^{4-5}$ supernovae over the last several million years to account for the
hot gas.  This is simply not plausible.  There is a series of OB associations and
recent star formation at the base of the NML in the vicinity of optical filament B, roughly 3.5$'$
(3.77 kpc in projection) from knot N4 \citep{graham98,mould00}.
They estimate that roughly 1.2$\times$10$^4$ M$_\odot$ of stars have formed in $\sim$16 Myrs.
This level of star formation is well below what would be required to create supernovae
driven bubbles on the scale seen in the Cen A NML.  There is no evidence for star formation on
such a massive scale as required to produce SN-driven winds on the observed
scale in the NML.

\begin{figure}
\plotone{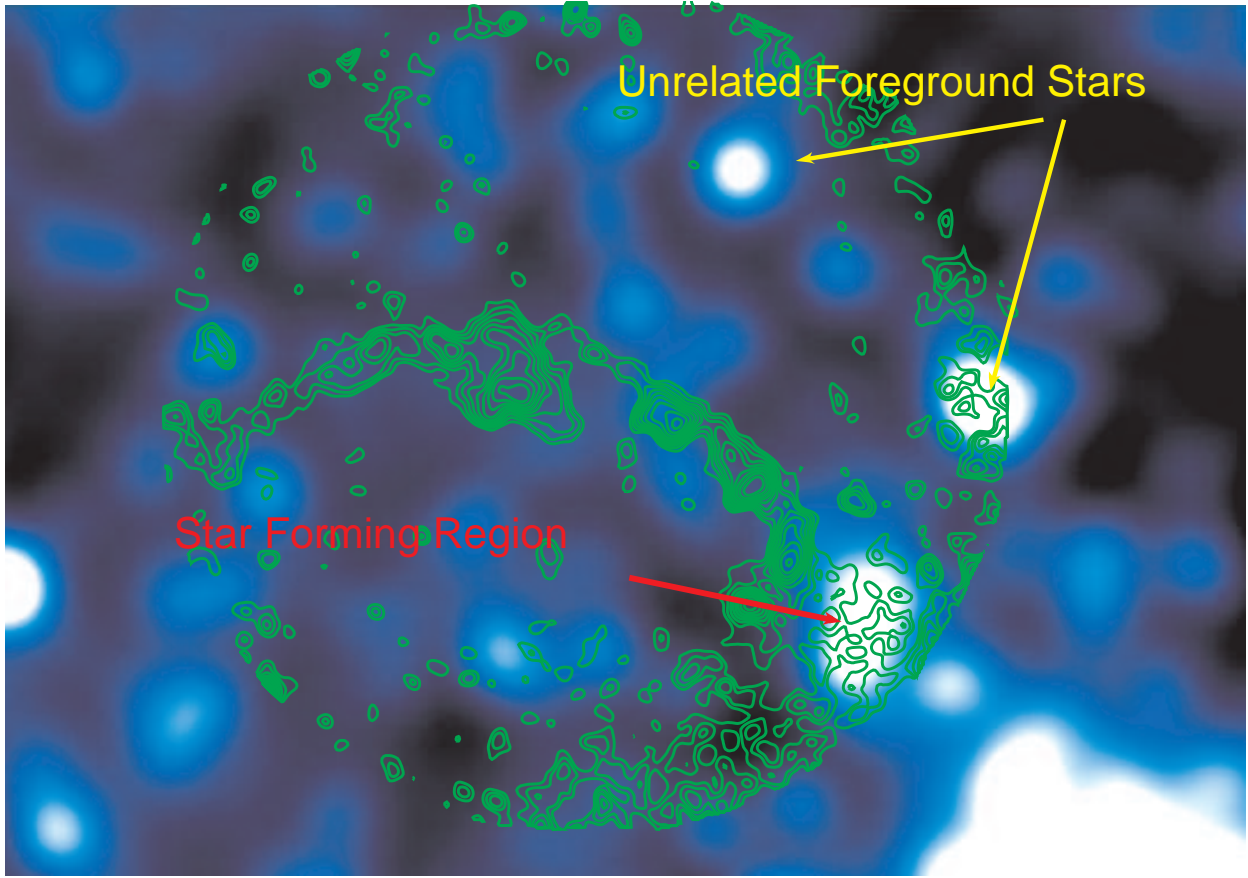}
\hspace{0.5in}
\caption{{\em XMM-Newton} contours (from Figure~\ref{nmlsmooth}) overlaid onto a
GALEX NUV image.  The point sources and all other features on spatial scales
smaller than 30$''$ arcseconds have been removed via wavelet decomposition.  Diffuse
emission from the well studied star forming region at the base of the NML
is shown with the red arrow.  Two bright foreground stars are also noted.}\label{galex}
\end{figure}

Contours from the X-ray image in Figure~\ref{nmlsmooth} have been overlaid onto an
archival GALEX NUV image of Cen A in Figure~\ref{galex}.
The position of the OB associations created by compression
of the HI cloud by the radio jet are clearly visible in the NUV GALEX image \citep{neff03}.
There are no other significant associations of hot stars associated with any
of the X-ray knots of the NML.  There does not appear to be any region of enhanced
star formation associated with the X-ray emitting gas.
We further processed the GALEX image using a wavelet
decomposition to remove all of the small-scale (i.e. point source) structure to search
for diffuse emission.  Again, there does not appear to be any significant enhancement of 
NUV emitting gas associated with the X-ray knots of the NML as there is for the star forming region
at the base of the NML.
Additionally, we examined archival Spitzer/MIPS observations of the NML region of Cen A
for any evidence of star formation or stellar clusters that may be related to the X-ray emission
and found none.  We therefore reject the possibility
that this gas is the result of a supernovae-driven wind.

\subsection{Photoionization}

It is possible that the X-ray filament was created by photo-ionization from the beamed
X-ray flux from the parsec scale jet at the nucleus.  However, the observed X-ray flux (i.e. into our line of
sight) from the accretion disk \citep{devans04} is orders of magnitude too small to create
X-ray knots 15-20 kpc away.  
We therefore require that there be a substantially larger flux of beamed EUV/X-ray emission
in the direction of the X-ray knots.  The spectral resolution of Si detectors
is not sufficient to distinguish between photoionized gas and collisionally excited
gas at X-ray temperatures.  We estimate the
required luminosity using arguments commonly invoked to explain the presence of X-ray emitting
gas in the vicinity of Seyfert nuclei.
If the gas is optically thin to soft X-rays (certainly true in our case), the ionization structure
is completely determined by the ionization parameter, $\xi$ (units of ergs cm s$^{-1}$), and is given by
$$\xi = \frac{L_{nucl}}{n_Hd_s^2},$$
where $L_{nucl}$ is the luminosity of the source of ionizing photons (ergs s$^{-1}$), $n_H$ is the
hydrogen density (cm$^{-3}$), and $d_s$ is the distance (cm) from the source to the X-ray emitting cloud.
The X-ray spectrum of each of the knots is dominated by emission from the Fe L complex
between 0.7 and 1.1 keV.  For the photoionization spectrum to be dominated by this emission,
$\xi\sim$100 \citep{kallman82, kallman91, weaver95}.
The X-ray luminosity of the knot, $L_{knot}$, is given by
$$L_{knot}=\frac{4\pi}{3} n_H^2 r^3 \Lambda(\xi),$$
where $r_{knot}$ is the radius of the knot and $\Lambda$ is the emissivity (ergs cm$^3$ s$^{-1}$)
of the photoionized gas.  For $\xi\sim$100, $\Lambda\sim$10$^{-24}$ ergs cm$^3$ s$^{-1}$ \citep{kallman82},
and we find $n_H\sim$ 4.2$\times$10$^{-2}$ cm$^{_3}$ (not surprisingly similar to
the collisional case).  

Using the observed values of X-ray flux for the brightest knot N1 ($F_x$=5.6$\times$10$^{-13}$
ergs cm$^{-2}$ s$^{-1}$ in the 0.1-10.0 keV bandpass (unabsorbed), $L_x$=9.1$\times$10$^{38}$
ergs s$^{-1}$) and at $\sim$30 kpc from the nucleus the most distant, we require
a nuclear luminosity of $\sim$6$\times$10$^{46}$ ergs s$^{-1}$, five orders of magnitude
larger than the observed nuclear flux \citep{devans04}.
In the unified scheme, Cen A is a misdirected BL Lac object \citep{morg91}, 
so that the X-ray flux within a few
degrees of the direction of the jet could be orders of magnitude larger than that we observe
from the nucleus.
This could be mitigated somewhat if the central source of photons is strongly beamed.
In fact, \citep{morg91} claimed that beamed flux from the nucleus could
explain the optical emission line filaments if the nuclear flux is $\sim$200 times larger
than the observed flux.
Large boosts can be obtained on-axis for moderate values of
$\Gamma$, but the five bright knots subtend an angle of $\sim$15$^\circ$
projected on the sky and are misaligned relative to the pc-scale jet and
the inner jet.  Assuming that the X-ray filament occupies a cone with an opening
half-angle of $\sim$7$^\circ$, the required value of $\Gamma$ is $\sim$8.
This value of $\Gamma$ would boost the flux from the nucleus toward the NML by a factor of roughly
250, thus dropping the beam power (on one side) to $\sim$2.4$\times$10$^{44}$ ergs s$^{-1}$.

There are at least five serious problems for this model.  First, this beamed flux would be
easily observable due to electron scattering from the hot gas in the central few
kpc of the galaxy.  The optical depth to electron scattering, $\tau_{es}$, is
given by $\tau_{es}=1.2n_H\ell \sigma_{es}\sim$10$^{-5}$.  The X-ray luminosity of the electron
scattered photons would be $\sim$5$\times$10$^{39}$ ergs s$^{-1}$ for the gas in the central
5 kpc of Cen A.  This would be easily observable (the integrated X-ray luminosity of all of the
hot gas in Cen A is only $\sim$10$^{40}$ ergs s$^{-1}$).
Second, we have chosen the smallest possible solid angle for the beamed flux.  The knots are
probably spread over a larger solid angle in three dimensions, and the pc-scale jet is misaligned
relative to the NML.  The luminosity quoted above is then only a lower limit.
Third, the luminosity and velocity of the jet are both uncomfortably large.  
We minimally require a beamed flux that is at least several times larger than that postulated by
\citet{morg92} to explain the inner optical filaments.
Fourth, there is no evidence for any interaction between the beamed flux from the nucleus
and cold/warm gas in the central regions of the galaxy, nor from any beamed flux on the counter-jet
side.  Such a large flux distributed over a fairly large solid angle
should leave some observable signature on the cold/warm gas (e.g. other photoionized clouds or perhaps
X-ray reflection nebulae similar to Sgr A*) in the central few kpc.
Fifth, if the X-ray filament of the NML were the result of photoionization, it would make
the close spatial relationship (in projection) to the radio lobe purely coincidental.
The photoionization model is implausible in our view.

\subsection{Supersonic motions of the external ISM}

As a third scenario, we consider the possibility that the X-ray filament
is created by the supersonic motion of the ambient ISM.  As described in
section 3.4, there is an extensive gas halo around Cen A extending at least
to the position of the NML.  Supersonic galaxy and cluster mergers have been
well studied by Chandra (the Bullet cluster being the best example).  Cen A is known
to have undergone a fairly recent merger (several hundred million years ago -
see the Introduction).  In this model, the X-ray filament represents
the shock front as the ISM moves supersonically to the NW.  The NML
could then have been re-energized by the shock.  The shock is not currently visible
in the NML as it would propagate rapidly through the lobe since the lobe's
sounds speed is high relative to that of the ISM.  There are several serious
problems with this model.  First, equating the ram pressure of the ambient gas
($1/2 \rho_{ISM}v_{ISM}^2$) with the pressure of the knots ($p_{knot}$ from Table 2) 
to estimate the gas velocity ($v_{ISM}$), we find that the
gas must be moving at $\sim$1700 km s$^{-1}$.  This is roughly Mach 6 for
the $k_BT$=0.3 keV ISM gas, implausibly large.  Second, even if the
merger took place at these velocities, it is unlikely that the velocity
remains supersonic several hundred years after the merger.
Additionally, there is no way that two small galaxies would be bound to each other
if they approached each other with this velocity (i.e. the small spiral would have shot
past Cen A, not fallen into the center).  Fourth, if the NML has been shock heated,
it would have to be in roughly pressure equilibrium with the filament, which
does not appear to be the case.  This scenario is highly implausible.

\subsection{Interaction with the Radio Lobe}

We next consider if the X-ray knots are produced by a direct interaction
between the radio plasma of the NML and gas in the vicinity of or interior to the NML.  We consider three
scenarios including shock heating of the ambient ISM by
the supersonic inflation of the lobe, entrainment by a buoyant bubble, 
and a direct interaction between the momentum of an
unseen jet with cold clouds embedded in the NML.  Before discussing each of these scenarios
in detail, we first present a re-analysis and discussion of the existing radio data on
the NML to constrain the relationship between the radio plasma and the hot gas.
We computed the equipartition pressure of the NML using a C-band radio map.
We assume that the NML is a cylinder of radius 6.5$'$ and length of 21.7$'$.
The flux density of the NML is 25.5 Jy at 5 GHz.  Using standard assumptions about
the low energy particle distribution (i.e. no protons and $\gamma_{min}$=100), 
we find an equipartition pressure of 1.2$\times$10$^{-12}$
dyn cm$^{-2}$ and an equipartition magnetic field, $B_{eq}$, of $\sim$7 $\mu$G.

Radio polarization maps could, in principle, constrain the relationship of the radio
emitting plasma to the X-ray emitting gas via measurement of the rotation measure.
The rotation measure map of the Cen A NML (Figure~5 of \citet{morg99}) shows systematic
variations of the rotation measure between +40 rad m$^{-2}$ and -160 rad m$^{-2}$.  There are
three arcminute-scale regions along the southeastern boundary of the lobe where the
rotation measure changes from -40 rad m$^{-2}$ to -160 rad m$^{-2}$ on scales of tens of
arcseconds.  None of these regions coincide with an X-ray knot, so unfortunately no
information about the position of the knots relative to the radio plasma can be drawn.
Additionally, the implied {\it minimum} magnetic field (assuming the kpc-scale features of the X-ray
knots associated with the NML and densities taken from Table~\ref{paramtab}) required to
create the observed rotation measure is on the order of 10 $\mu$G.  
This magnetic field is not unrealistic, although somewhat larger than the
few $\mu$G fields one might naively expect.  The pressure of the knots would still be
dominated by the hot gas.  It is in our view more
likely that unrelated gas on much larger spatial scales (with correspondingly lower magnetic field and
lower density) is responsible for the observed variations in rotation measure. 
\citet{morg99} point out that the X-ray filament is associated with the region of
lowest fractional polarization, suggesting that the hot gas is either a screen in front
of the lobe or intermixed with the lobe plasma.

\subsubsection{Supersonic inflation of the NML}

In the scenario where that the X-ray filament has been created by the
supersonic inflation of the NML, the NML is still being powered by
the active nucleus and cold or warm gas has been shock-heated to keV temperatures.
The NML is then an energy-driven bubble driving a shock into the ambient gas.  Such a model
has been invoked to explain the hot shell of gas around the SW inner lobe of Cen A \citep{kraft03},
a large number of shocks seen in galaxy groups and clusters of 
galaxies \citep{forman05,nulsen04,mcnamara05}, and hard X-ray emission associated with
CSS radio source 3C 303.1 \citep{odea06}.
In this case, the knots/bridge of the NML were originally warm/cold gas in rough hydrostatic
equilibrium with the hot ISM that was shock heated by the NML.  There are several
attractive features of this scenario.  First, the sound crossing time of the knots
is roughly consistent with the age of the shock around the SW radio lobe (roughly 3 Myrs, 
see \citet{kraft03,croston09}).  If this age
also represents the approximate age of the NML, the
NML and the SW (inner) lobe are part of the same outburst.
The momentum imparted to dense clouds in a stratified atmosphere by the passage of a strong
shock is small, and the sites of the knots could represent the positions of dense clouds
that have been shock heated.  The tenuous filament between the knots then represents
material that has been ablated from the dense clouds.

The main problem with this scenario is that the equipartition pressure of the
radio lobe is more than an order of magnitude below that of the knots.  
This is, of itself, not a serious problem.  In fact, we have shown that the SW
inner lobe must be far from equipartition and greatly overpressurized relative to the
ambient medium \citep{kraft03,croston09}.  In addition, {\em Chandra} and
{\em XMM-Newton} observations of the hot gas atmospheres surrounding radio lobes show that
the equipartition pressure of lobes can be more than an order of magnitude less than that
of the ambient pressure, particularly for FR I sources.  It is difficult to see
how the pressure of the NML can be substantially less than that of shock-heated features
which the lobe has already overrun (i.e. they are internal to the lobe).  Additionally,
at the equipartition pressure, the NML is currently in rough pressure equilibrium with the
ambient ISM (based on an extrapolation of the surface brightness profile of the ISM derived from
the {\em Chandra} image to the distance of the NML).  Increasing the equipartition energy and pressure
of the lobe to roughly match that of the X-ray knots
would then imply the NML is enormously overpressurized relative to the ambient
ISM.  Additionally, the total energy of the NML, at the equipartition value, is roughly the
same as that of the SW (inner) lobe.
If the NML were in rough pressure equilibrium with the knots, the internal energy of the NML
would be more than an order of magnitude larger than the SW inner lobe.  Thus the knots
and the SW lobe could not be
from the same outburst, but the NML could also not be the result of an earlier epoch of
nuclear activity given its short lifetime.  This scenario leads to a contradiction.
Additionally, the lack of any counterpart to the NML on the southwestern side of Cen A becomes even
more puzzling.

\subsubsection{Entrainment}

A fourth scenario is that the X-ray filament consists of material that
is entrained by the NML as it is driven or rises buoyantly in the hot gas atmosphere \citep{sax03}.
The entropy of the gas in the knots is roughly consistent with that in
the center of Cen A, and we typically expect only adiabatic compression
if the gas is entrained via Kelvin-Helmholtz instabilities in the linear regime.
There are several serious problems with this interpretation, however.
Most importantly, the short lifetimes and tremendous overpressurization of the X-ray knots
relative to the NML make the model of buoyant entrainment untenable.
The equipartition magnetic field of the radio lobe is $\sim$7 $\mu$G,
the equipartition pressure of the lobe is $\sim$1.2$\times$10$^{-12}$ dyn cm$^{-2}$.
This is more than an order of magnitude lower than the thermal pressures of the
various knots.  Dissipation and heating will invariably occur in gas that has been entrained via
Kelvin-Helmholtz instabilities, but it is difficult to see how such large pressure
gradients could be formed and stably maintained.
The buoyant rise time of the NML from the nucleus is
$\sim$170 Myrs \citep{sax03}.  If the X-ray knots are unconfined, they will expand at
roughly their internal sound speed until they reach pressure equilibrium with the lobe
(or other ambient gas that dominates the pressure).  The sound crossing time of the
X-ray knots is more than an order of magnitude smaller than the buoyant rise time
to the position of the NML.  The only way that this scenario is plausible
is if the NML is far from equipartition.

\subsubsection{Cold gas directly impacted by the jet}

The remaining viable scenario to explain the X-ray filament is that
the X-ray knots and bridge are (or were) directly embedded in the jet.  
That is, cold, dense clouds lie directly in the
path of the jet and are in direct contact with the jet momentum.
This is the only scenario that can reasonably reproduce the observed features of the X-ray filament
without requiring unrealistically large input energies or inconsistencies
related to timescales.  The NML expands to the NW of the X-ray filament, so it is
reasonable to speculate that the jet was deflected to the NW by the interaction with the
cold gas (i.e. the collision that created the X-ray filament was oblique with the jet generally
lying to the NW of the cold gas).

Simulations of jet-cloud interactions demonstrate that a dense obstacle can
deflect a jet \citep{deyoung91, deyoung02, higgins99, wang00}.
In fact, many features of these simulations qualitatively reproduce structures seen in
the X-ray filament of the NML.  Strong shocks will be driven into the dense clouds.
The X-ray knots of the NML could represent denser clouds of gas that have been shock heated by
an oblique interaction with the jet.  The filamentary bridge structure between the knots is then
gas that has been heated and ablated from the dense clouds.
Simulations of the propagation of a supersonic jet through a stratified medium consisting
of a hot, low density gas that surrounds cooler, denser clouds show the underlying flow patterns
can be quite complex in the case in which the filling factor of the cold gas is large \citep{saxton05}.
The momentum of the jet tends to propagate through a number of small channels as the
dense clouds are shock-heated and subsequently ablated.  Ablation and shock-heating of gas
is entirely consistent with our data; each of the bright knots could be a cloud and
the bridge connecting the knots could be ablated gas that has been swept along the direction of
the flow.  This scenario naturally explains the observed anti-correlation of the X-ray and
radio knots of the NML.  The radio plasma does not penetrate the clouds but is compressed
into the channels between them with a significant increase in their radio brightness due to
compression of the magnetic field.

This model requires that the NML is actively powered by
the central engine, but whether it is part of the current outburst or a re-energized bubble from
a previous outburst is not clear.  As we describe below, we favor the latter scenario.
It also requires that the NE inner lobe is really a `jet' in the sense that it
acts as a nearly lossless tunnel through which the collimated material from the nucleus flows
out to the NML.  The term `NE inner lobe' is therefore 
somewhat of a misnomer, it is really a jet with at least two
bends in it:  the first at the transition between the inner jet and inner lobe (where the flow
must become subsonic relative to its internal sound speed), and a second as
the inner lobe bends into the large scale radio jet.  
Subsonic or transonic gas motions of the external medium
are likely the cause of the jet bends \citep{kraft08}.
We also note that we implicitly assume in the discussion that follows that the jet mechanical power
is equal on either side of the AGN (both instantaneously and in a time averaged sense over a long
period).  If the jet power/momentum were unbalanced on the two sides of the AGN, the central SMBH would be kicked out
of the central region of the galaxy on a short timescale.  This is a conservative
assumption as there are no known examples of this process occurring.

The energetics of the NML and the SW inner lobe are surprisingly similar, strengthening
the argument that they are both related to the current outburst.  Assuming the geometry
for the NML used in the equipartition argument above and the equipartition pressure, 
the enthalpy (for $\gamma$=4/3) of the NML is $\sim$3$\times$10$^{56}$ ergs.
This is really a lower limit to the total enthalpy, but as we argued above, the NML is
probably not far from equipartition.  The pressure of the SW inner lobe is
$\sim$1.6$\times$10$^{-11}$ dyn cm$^{-2}$ \citep{kraft03,croston09}.  Assuming a
simple elliptical geometry for this lobe, its enthalpy is $\sim$6$\times$10$^{56}$ ergs.
Roughly half the energy that has gone into inflating the SW inner lobe is required to
inflate the NML.  Additionally, the total thermal energy of the five knots in the NML is
$\sim$2.6$\times$10$^{55}$ ergs.  Thus only five percent of the energy required to inflate
the NML must go into shock-heating cold gas to create the X-ray knots.
Based on energetics, it is entirely plausible that the NML and the SW inner lobe are part of the outburst,
and only a small fraction of the outburst energy must be diverted to shock-heating the X-ray
emitting knots.

\begin{figure}
\plotone{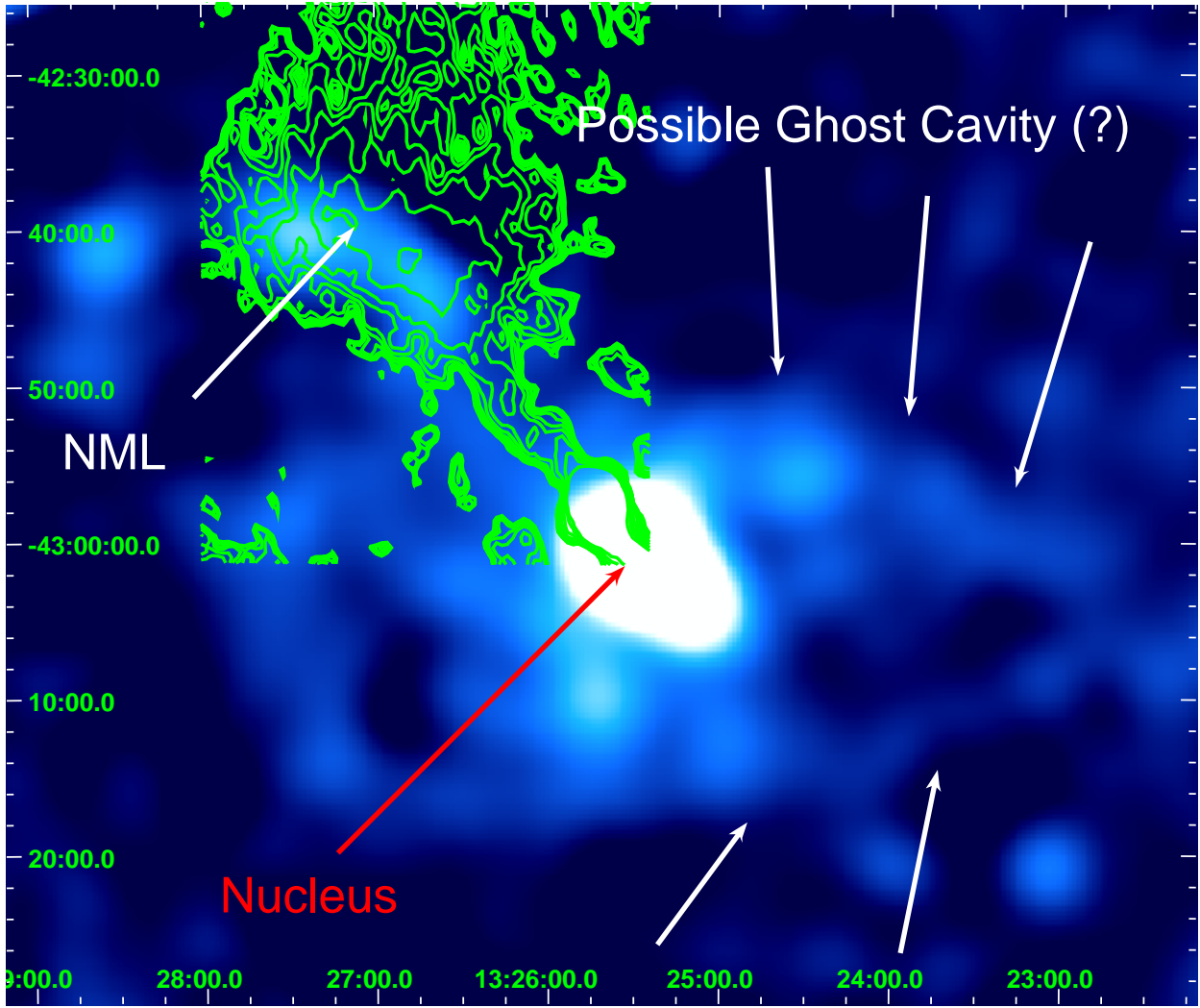}
\hspace{0.5in}
\caption{Exposure corrected {\em ROSAT} PSPC observation (14 ks) of Cen A in the 0.5-1.1 keV bandpass.  The
radio contours of the NML are overlaid, and the possible `ghost' cavity is denoted by the
white arrows.}\label{rosat}
\end{figure}

The problem with associating the NML directly with the current outburst is that the average
inflation speed of the bubble must have been considerably higher than the expansion
velocity of the SW inner lobe given that the NML is both further away from the nucleus and
considerably larger than the SW inner lobe.  At minimum pressure the NML appears to be in rough pressure equilibrium
with the ambient gas, and there is no evidence for a shell of hot gas around the NML that would
be expected if it had recently driven a shock.  Thus, it is more likely that the NML
does indeed represent a previous epoch of nuclear activity.  A direct channel between the
NML and NE inner lobe has been opened and the NML is now being powered.
This requires that there is an unseen radio bubble lying several tens of kpc to the SW of the active
nucleus of Cen A that was created in this earlier epoch of nuclear activity.
Non-detection of a radio bubble at hundreds of MHz and/or GHz frequencies does not
conclusively rule out the existence of a radio bubble.
{\em Chandra} observations of NGC 4636 show clear evidence of shocks being driven into
the ISM by unseen radio bubbles/jets \citep{jones02, baldi08}.
Ghost cavities have also been observed in a number of objects in which there are clear
depressions in the X-ray emission of group/cluster gas associated with the inflation of radio
lobes but faint or undetected radio bubbles \citep{mcnamara01,choi04}.
Radiative or adiabatic losses could easily account for the near invisibility of
a putative southwestern counterpart to the NML.  Interestingly a short (14 ks) archival {\em ROSAT} PSPC
observation of Cen A shows marginal evidence for a ghost cavity to the southwest of
the nucleus.  A smoothed, exposure corrected {\em ROSAT} PSPC image of Cen A in the 0.5-1.1 keV
band is shown in Figure~\ref{rosat}.  The white arrows delineate what may be a ghost
cavity.  If the NML is the result of a previous outburst, our conclusions are largely
unchanged.  The energy required to re-energize GHz emitting particles of the NML is obviously
less than the total enthalpy of the lobe, requiring an even smaller fraction of the energy of the current outburst.

One obvious question that presents itself in the jet-cloud model is that of
where the cold gas originated.
We consider three possible scenarios.  First, we consider a model in which
a direct interaction between the jet and the HI cloud at the base of the NML
imparted sufficient momentum to drive the knots out to their current position.
Again we consider knot N1 as the most distant and therefore demanding
case.  The mass of this feature is $\sim$3.5$\times$10$^6$ $M_\odot$ and it lies roughly 10 kpc from
the HI cloud.  Its lifetime is $\sim$ 3 Myrs, so that is must have a velocity of
$\sim$3000 km s$^{-1}$ to reach its position relative to the HI cloud.  The acceleration
required to achieve this velocity in this time is $\sim$3$\times$10$^{-6}$ cm s$^{-2}$.
If we assume that the jet is relativistic (i.e. light), the mechanical energy required to accelerate knot N1
to 3000 km s$^{-1}$ in 3 Myrs is $\sim$1.2$\times$10$^{59}$ ergs or a mechanical power of
$\sim$1.3$\times$10$^{45}$ ergs s$^{-1}$.  This is two orders of magnitude larger than
the mechanical power of the SW lobe, and is in any case more typical of FR II radio galaxies.  The energy
and momentum required to push cold gas from the HI cloud to knot N1 is implausibly large, and
we conclude that most of the gas in the X-ray filament did not originate in this HI cloud.  
On the other hand, we could assume that the jet is heavily mass-loaded.  It may have started as a relativistic
plasma, but has swept up considerable mass (perhaps via Kelvin-Helmholtz instabilities along the
jet/ISM interface or from stellar mass loss of stars embedded in the jet).
The energy requirements are then greatly reduced, but the jet density required is comparable to
or greater than the density of the ambient ISM and thus implausibly high.
The X-ray knots must have been shock-heated in situ near to their present positions.

Second, the knots could have recently drifted into the jet.  The edges of
the clouds are then continuously torn off as they propagate into the beam.  Given the age
of the knots (3$\times$10$^6$ yrs) and assuming they are rotating in the gravitational
potential of the galaxy, they could only have travelled $\sim$1 kpc in the lifetime of the jet (i.e
just outside the jet).  This scenario would naturally explain the bridge of emission between the knots
as it represents material ablated by the jet.  This model also seems unlikely as it
requires a chain of knots originally located just outside the NML to drift in.
A third scenario (and most likely in our view) would be that the clouds
drifted into the NML prior to the current outburst and are now being lit up by the
current epoch of re-energization of the NML.  Transonic or subsonic gas motions of the external medium
can easily push (and deform) the NML over a timescale of hundreds of megayears.  

In this jet-cloud interaction scenario, the jet must survive two abrupt bends to provide 
a collimated outflow to the NML.  Light, highly supersonic jets cannot be bent
without typically disrupting the jet on a short timescale, demonstrating that the
jet must be subsonic at the end of the inner jet.  Hydrodynamic modeling
of mass loading of the jet, assuming it is pressure confined by the external medium,
shows that the jet likely decelerates to subsonic velocities just as it enters
the NE inner lobe (P. Nulsen \etal~in preparation).
Once the jet becomes subsonic, external subsonic or transonic gas motions can easily
deform the jet.  Based on a deep {\em Chandra} observation, we have already shown the existence of
a surface brightness discontinuity in the gas at the base of the NE lobe and speculated
that the NE inner lobe may have inflated in a cross-wind \citep{kraft08}.  
Observationally, there are a large number of examples of jets remaining
well collimated after a bend on scale of pcs (e.g. VLBI observations of BL Lac jets)
to tens of kpc (e.g. Her A).  
The bending of jets in wide and narrow angle tailed galaxies observed in clusters of galaxies are believed
to be the result of ram pressure bending the jets \citep{pinkney94}, although these jets
typically do not remain well collimated.
Once a stable connection between the NE inner lobe and the NML is formed, energy can
be rapidly transferred to the NML as the velocity of the jet, while subsonic, is still likely
to be considerably larger than any large scale dynamical motions external to the jet.
The large scale jet in Cen A connecting the NE inner lobe to the NML shows no evidence
of decollimation.  In fact, the opening angle of the LSJ appears to be constant between the NE
inner lobe and the NML, qualitatively consistent with the large scale uniform density gas halo.

Additionally, it is not easy to understand how the bridge of X-ray emission extends
$\sim$ 5 kpc in projection beyond knot N1 and beyond the radio contours of the NML.
By the same argument above, transonic motions of the external medium can only push the
NML $\sim$1 kpc in 3$\times$10$^6$ yrs.  The actual boundaries of the NML to the northeast and
northwest are not well defined in the radio maps.  
The NML opens into much larger scale radio emission extending hundreds of kpc from the nucleus without
any clearly defined interface.  We can only speculate that
the extension of the X-ray filament beyond knot N1
and the observed radio contours of the NML implies that the jet extends beyond the visible radio 
emission bending east from knot N1 (i.e. the radio emission shown on Figures~\ref{nmlovl} and~\ref{nmlgap}
from the NML does not entirely define the location of the jet).

Finally as a minimal consistency check, we require that the pressure of the individual knots
be on the order of or less than the ram pressure of the jet, or $p_{knot}< 2P_{jet}/v_jA$ where
$p_{knot}$ is the pressure of a knot, $P_{jet}$ is the jet power, $A$ is the cross-sectional
area of the knot, and $v_j$ is the jet velocity.  In particular, we require that the
upper limit to the speed of the jet is much larger than the thermal 
speed of the clouds so that the timescale for the jet interaction is smaller than any
dynamical timescale for a cloud.  From Tables~\ref{specfit} and~\ref{paramtab},
the pressure of a typical cloud is $\sim$2$\times$10$^{-11}$ dyn cm$^{-11}$ with radius
$\sim$1 kpc.  We estimate the power of the jet (6$\times$10$^{42}$ ergs s$^{-1}$)
based on the enthalpy of the SW inner lobe
and the velocity of the shock \citep{kraft03,croston09}.
For these values of the knot pressure, knot cross-section, and jet power, we find that
$v<0.7c$.  Even if only a fraction ($<$10\%) of the jet power is intercepted by the X-ray knots,
the jet velocity can easily be thousands or tens of thousands of km s$^{-1}$,
well above any thermal velocity of any of the knots.

\section{Summary and Conclusions}

We present results from a 40 ks {\em XMM-Newton} observation of an X-ray filament associated with the Cen A
Northern Middle Radio Lobe.  Our conclusions can be summarized as follows:
\begin{enumerate}
\item The filament is composed of at least five discrete knots of emission connected by a
narrow thread that lies along the southeast boundary of the radio lobe.
\item The spectra of all the knots and the connecting thread are well described by thermal
models with low abundance, thus confirming that the emission is from hot gas.
\item We also report the discovery of a large scale X-ray emitting corona around Cen A.
\item Our preferred model for the origin of this X-ray filament is a jet-cloud interaction.
Cold, dense clouds have been shock heated to X-ray temperatures. 
\item It is likely that the NML was created by a previous outburst,
but it is currently being powered by the ongoing nuclear outburst if our preferred
model for the X-ray filament is correct.
\end{enumerate}

The offsets between the optical emission-line gas and the X-ray emitting gas in the Cen A
NML have important implications for our understanding of both the EELR in other radio
galaxies (see references above) and the interaction of the relatively weak jets of Seyfert galaxies with the
the ambient multiphase gas of the host galaxy \citep{devans06, croston08}.  In particular, the spatial relationship 
between the radio jet, the X-ray emitting gas, and the warm optical emission-line gas is often not clear in these more
distant galaxies.  In the Cen A NML, we find that the optical emission-line gas is not
co-spatial and intermixed with the X-ray emitting gas.
However the jet is interacting with the ambient gas to create the observed filaments,
the warm gas is clearly separated from the X-ray emitting gas even though both are
greatly overpressurized relative ambient coronal gas and therefore short-lived.

There are at least four further observational programs that could be undertaken to confirm
the results and interpretations presented here.  First, a deep study of the large-scale
X-ray corona would confirm or reject the presence of the ghost cavity hinted at in the
{\em ROSAT} observation.  A detailed study of the morphology and thermodynamic parameters of the
gas in the vicinity of this cavity could constrain its age and total energy and would
provide an important comparison of the results presented here.  Second, there may well
be considerably more HI around Cen A than that detected by previous studies.
HI observations of early-type galaxies around the periphery of the Virgo cluster show
little evidence for HI, but the ambient gas density is high enough that ram pressure stripping
may be important in removing HI from galaxy halos.  Cen A is in an isolated environment.
so primordial HI may well survive into the present epoch.  \citet{sch94} find 1.5$\times$10$^8$ $M_\odot$
of HI in a shell out to 15$'$ from the nucleus.  The X-ray filament lies beyond this shell,
but the filament mass is small compared to the known HI.  Scaling by the solid angle subtended
by the NML, there could easily be an additional $\sim$10$^8$ $M_\odot$ of HI more than
15$'$ from the nucleus.  The existing HI survey of the Cen A group \citep{banks99} is not sensitive
to the relatively small amounts of cold gas that are in the NML.  A deeper, wider HI survey is required
to evaluate the possible existence of this cold gas.  
Fourth, a more sensitive, higher resolution radio polarization study combined with the
XMM-Newton data presented here, could conclusively determine the spatial relationship of the
hot gas and the radio plasma of the NML.
Finally, deep {\em Chandra} observations
of the individual knots will give unique insights into the dynamics of the jet-cloud
interaction.  The unphysically low values found for the elemental abundances in the spectral
fits suggest the presence of unresolved temperature structure in each of the knots.
This structure would only be resolved with {\em Chandra} (on-axis), facilitating a unique
comparison with hydrodynamic simulations of jet-cloud interactions.

Finally we note that this phenomenon would be nearly unobservable in most other radio galaxies given
the relatively low luminosity of the X-ray knots.
It is possible that the jet-cloud interaction is a relatively common occurrence, and that
we haven't observed many such examples simply due to a lack of sensitivity.
If Cen A were 20 times futher away (70 Mpc), only a few tens of counts would be detected
from the knots and none of the substructure would be observable.  
Because of the distance of the X-ray filament from the nucleus (15-30 kpc) and the inability
to distinguish thermal from non-thermal models with such few counts, it is likely
that the X-ray emission would be attributed to the synchrotron process.
Additionally, Cen A is a relatively low power radio galaxy compared
with many of the more distant 3C radio galaxies.  In the more powerful radio galaxies,
the jets are likely to remain light and fast to fairly large distances.  In Cen A, the
fact that the jet apparently bends (twice) between the inner jet and the NML suggests
that the jet is transonic or subsonic and may therefore have swept up considerable mass.
The jet impacting the clouds in the NML may therefore be both slower and denser than
the jets of more powerful radio galaxies.  Any jet/cloud interaction in the latter category
of sources may simply result in the rapid destruction and incorporation of the cloud into
the jet as it is shock-heated to temperatures well above the soft X-ray band.

\section{Acknowledgments}

This work was supported by NASA contract NAS8-03060, the {\em Chandra} X-ray Center,
and the Smithsonian Astrophysical Observatory.  MJH thanks the Royal Society for a research fellowship.

\clearpage

\clearpage

\clearpage

\clearpage

\clearpage

\clearpage

\clearpage

\clearpage 

\clearpage

\clearpage

\clearpage

\begin{table}
\begin{center}
\begin{tabular}{|c|c|c|c|c|c|c||c|}\hline Region & RA & Dec & Radius & Temperature (keV) & $Z$($\times$ 10$^{-2}$) & $L_X$ & Norm ($\times$10$^{-5}$) \\ \hline\hline 
N1 & 13:27:25.3 & -42:40:17.5 & 1.48$'$ & 0.71$\pm$0.02 & 9$\pm$2 & 5.1 & 21. \\ \hline 
N2 & 13:26:56.8 & -42:41:37.4 & 0.98$'$ & 0.50$\pm$0.04 & 5$\pm$2 & 1.8 & 7.5 \\ \hline
N3 & 13:26:40.5 & -42:43:50.6 & 0.89$'$ & 0.58$\pm$0.07 & 9$\pm$2 & 2.0 & 8.1 \\ \hline
\multicolumn{4}{|c|}{} & 0.65$\pm$0.05 & 13$\pm$8 & & \\ \hline
N4 & 13:26:34.2 & -42:46:19.8 & 0.98$'$ & 0.39$\pm$0.03 & 5$\pm$3 & 1.8 & 10. \\ \hline
\multicolumn{4}{|c|}{} & 0.49$\pm$0.08 & $<$0.2 & & \\ \hline
N5 & 13:26:41.7 & -42:48:06.6 & 0.30$'$ & 1.07$\pm$0.02 &  $<$0.1 & 1.0 & 8.3 \\ \hline
\multicolumn{4}{|c|}{} & 4.1$^{4.7}_{1.5}$ & $<$1.0 & & \\ \hline
\end{tabular}
\end{center}
\caption{Best fit temperatures and uncertainties (90\% confidence)
for five knots of the X-ray filament of the Cen A NML.  The regions with the
$C$ suffix are fits to the {\em Chandra} data.  Uncertainties are 90\% confidence for
one parameter of interest, and the coordinates are J2000.  The second set of temperatures
and elemental abundances for N3, N4, and N5 are from the {\em Chandra} data.  The $L_X$ column
is the X-ray luminosity of the knot in the 0.5-2.0 keV band (unabsorbed) in units of
10$^{38}$ ergs s$^{-1}$.  The norm column
contains the XSPEC normalization for the spectral fit for $Z$=0.5 (fixed) which
are used to derive the densities, pressure, energies, and masses of the knots
shown in Table~\ref{paramtab}.  These are not the normalizations for the best-fit, low abundance spectra.}\label{specfit}
\end{table}

\clearpage

\begin{table}
\begin{center}
\begin{tabular}{|c|c|c|c|c|c|}\hline
Region & Density ($n_H$) & Pressure & Mass & Energy & Lifetime \\
     & (10$^{-2}$ cm$^{-3}$) & (10$^{-11}$ dyn cm$^{-2}$) & (10$^6$ M$_\odot$) & (10$^{54}$ ergs) & (Myrs) \\ \hline\hline
  N1 & 0.8 &  2.2 & 3.5 & 16.5 & 3.7\\ \hline
  N2 & 1.1 &  2.0 & 1.0 & 3.2 & 2.6 \\ \hline
  N3 & 1.1 &  2.4 & 1.0 & 3.8 & 2.4 \\ \hline
  N4 & 1.2 &  1.6 & 1.1 & 2.6 & 3.1 \\ \hline
  N5 & 2.7 & 23.8 & 0.4 & 6.7 & 0.6 \\ \hline\hline
Lobe &     & 0.12 &     & 300 &     \\ \hline
\end{tabular}
\end{center}
\caption{Thermodynamic parameters of the knots in the Cen A NML X-ray filament assuming $Z$=0.5.  The
XSPEC normalizations from Table~\ref{specfit} have been scaled to $Z$=0.5 thus reducing the gas
density, pressure, and total mass as described in the text.  The values listed as pressure and
energy for the lobe are the equipartition pressure and the bubble enthalpy (for $\gamma$=4/3 assuming
equipartition) of the lobe.}\label{paramtab}
\end{table}

\end{document}